\documentclass{elsart}

\usepackage{natbib}

 \usepackage{graphicx}
 \usepackage{epsfig}

\usepackage{amssymb}

\begin{document}

\begin{frontmatter}



\title{A Unified Model of Short and Long Gamma-Ray Bursts, 
X-Ray Rich Gamma-Ray Bursts, and X-Ray Flashes}


%

%
%
%

\author{Ryo~Yamazaki}
\address{Department of Earth and Space Science, 
Graduate School of Science, Osaka University, Toyonaka, Osaka 560-0043, Japan}
\ead{ryo@vega.ess.sci.osaka-u.ac.jp}

\author{Kunihito~Ioka}
\address{Physics Department and Center for Gravitational Wave Physics,
104 Davey Laboratory, Pennsylvania State University, University Park,
PA 16802, USA}

\author{Takashi~Nakamura, Kenji~Toma}
\address{Theoretical Astrophysics Group
Department of Physics, Kyoto University
Kyoto 606-8502, Japan}

\begin{abstract}
We propose a possible unified model of short and long gamma-ray bursts
(GRBs), X-ray rich GRBs, and X-ray flashes.
The jet of a GRB is assumed to consist of multiple sub-jets or 
sub-shells (i.e., an inhomogeneous jet model).
The multiplicity of the sub-jets along a line of sight $n_s$ is 
an important parameter. 
If $n_s$ is large ($\gg 1$) the event looks like a long GRB, 
while if $n_s=1$, the event looks like a short GRB. 
Finally, when $n_s=0$, 
the event looks like an X-ray flash or an X-ray rich GRB.
Furthermore, our model may also explain the
 bimodal distributions of $T_{90}$ duration of BATSE-GRBs.
\end{abstract}


\end{frontmatter}

\newcommand{\tot}{{\rm tot}}
\def\N{\nonumber}
\def\f{\frac}
\def\max{{\rm max}}
\def\min{{\rm min}}
\def\j{{\scriptscriptstyle (j)}}
\def\tot{{{\rm tot}}}
\def\sub{{{\rm sub}}}
\def\obs{{{\rm obs}}}
\def\dep{{{\rm dep}}}


\section{Introduction}
While the association with the supernova is almost established or strongly
suggested for the long gamma-ray bursts (GRBs) \citep{stanek2003,hjorth03},
the origins of short GRBs and X-ray flashes (XRFs) remain unclear.
The observed event rate of  short GRBs is about a third of the long GRBs 
while the observed event rate of  XRFs is also about a third.
Although there may be a possible bias effect to these statistics, 
these numbers are, in an astrophysical sense, the same or  comparable. 
If these  phenomena arise from essentially different origins, 
the similar number of events is just by chance. 
On the other hand, if they are related, 
the similar number of  events is natural and 
the ratio of the event rate tells us something about
the geometry of the central engine.
We propose a unified model in which
the central engine of short GRBs, long GRBs and  XRFs is the same  
and the apparent differences come essentially from  different viewing
angles. For details, see \citet{yin04b} and \citet{toma2004}.

\section{Unified Model}
Roughly speaking,  short GRBs are similar to 
the first 1 sec of long GRBs \citep{ggc03},
which suggests that the difference between  short 
and long GRBs is just the number of pulses, 
and each pulse is essentially the same.
Thus, we may consider that each pulse is produced
by essentially the same unit or the sub-jet, and the GRB jet consists 
of many sub-jets.
If many sub-jets point to our line of sight, the event looks like 
the long GRB while  if a single sub-jet points to us, 
the event looks like a short GRB.
Since we can observe only the angular size of $\sim \gamma^{-1}$
within the GRB jet with the Lorentz factor $\gamma$,
different observers will see different number of sub-jets
depending on the distribution of sub-jets within the GRB jet.

XRFs also appear to be related to GRBs.
Softer and dimmer GRBs smoothly extend to the XRFs
\citep{He01a,ki02,s03},
Other properties of the XRFs are also similar to those of the GRBs,
suggesting that XRFs are in fact soft and dim GRBs.
In the sub-jet model, XRFs are naturally expected 
when our line of sight is off-axis to any sub-jets
\citep{nakamura2000,in01,yin02,yin03b,yyn03,yin04a}.

\begin{figure}
\begin{center}
\includegraphics*[width=14cm]{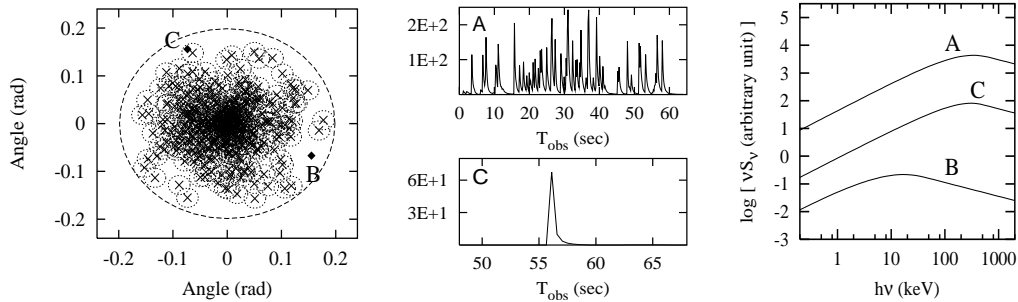}
\end{center}
\caption{
(Left)
The angular distribution of 350 sub-jets confined 
in the whole GRB jet in our simulation.
The whole jet has the opening half-angle of 
$\Delta\theta_\tot=0.2$~rad.
The sub-jets have the same properties.
The axes and the angular size of sub-jets are represented by crosses
and the dotted circles, respectively. 
 ``A'' represents the center of the whole jet and is hidden by the lines
 of sub-jets. 
(Middle)
The observed  $\gamma$-ray light curves,
corresponding the cases ``A''and ``C'' in the left panel.
(Right)
Time-integrated energy spectrum of the emission from the multiple sub-jets
for the observers denoted by 
``A'', ``B'',  and ``C''.
The source is located at $z=1$. 
}
\label{figs}
\end{figure}

In the following, we show a numerical simulation to demonstrate how
the event looks so different depending on the viewing angle 
in our unified model \citep{yin04b}.
The whole jet, with an opening half-angle of
$\Delta\theta_\tot\sim0.2$~rad, consists of 350 sub-jets.
All sub-jets are assumed to have the same intrinsic
properties, e.g., the opening half-angle
$\Delta\theta_\sub^\j=0.02$~rad, and so on.
We consider the case in which 
the angular distribution of  sub-jets is given by 
$P(\vartheta^\j)\propto
\exp[-(\vartheta^\j/\vartheta_c)^2/2]$,
where we adopt  $\vartheta_c=0.1$~rad \citep{z03}.
In this case, sub-jets are concentrated on the central region.
For our adopted parameters,  
isolated sub-jets exist near the edge of the whole jet 
with the multiplicity $n_s= 1$
and there exists a viewing angle where no sub-jets are launched.
The left, middle, and right panels of Fig.~\ref{figs} show 
the angular distributions of sub-jets and
the directions of three selected lines of sight, 
the observed light curves in the $\gamma$-ray bands, 
and the observed time-integrated spectra, 
respectively. Note here in the left panel of Fig.~\ref{figs}, 
``A'' represents 
the center of the whole jet and is hidden by the lines of sub-jets. 

{\it Long GRB}: 
When we observe the source from the $\vartheta=0$ axis
(case~``A''), we see spiky temporal structures
(the upper-middle panel of Fig.~1)  and  $E_p \sim 300$~keV
which are typical for the long GRBs.
We may identify  case ``A'' as  long GRBs. 

{\it XRF and X-ray rich GRB}: 
When the line of sight is away from any sub-jets
(case~``B''),
soft and dim prompt emission, i.e.  XRFs
or X-ray rich GRBs are observed with $E_p= 10\sim 20$~keV and $\sim 4$
orders of magnitude smaller fluence than that of   case ``A''
(the right panel of Fig.~1). 
The burst duration is comparable to that in  case  ``A''. 
These are quite similar to the characteristics of XRFs. 
We may identify the
case ~``B'' as  XRFs or X-ray rich GRBs.

{\it Short GRB}: 
If the line of sight is  inside an isolated sub-jet (case~``C''),
its observed pulse duration is $\sim 50$ times smaller than 
 case ``A''.
Contributions to the observed light curve from the other sub-jets
are negligible, so that the fluence is  about a hundredth
 of the case ``A''. 
These are quite similar to the characteristics of  short GRBs.
However the hardness ratio 
($=S(100-300~{\rm keV})/S(50-100~{\rm keV})$)  is about 3 
which is smaller than the mean hardness of short GRBs ($\sim 6$). 
\citet{ggc03} suggested that the hardness of  short GRBs
is due to the large low-energy photon index $\alpha_B\sim -0.58$ 
so that if the central engine
launches $\alpha_B\sim -0.58$ sub-jets to the periphery of the core 
where
 $n_s$ is small, we may identify the case ``C'' as the short-hard GRBs.
In other words, the hardness of 3 comes from $\alpha_B =-1$ in our simulation
so that if $\alpha_B\sim -0.58$, the hardness will be 6 or so. 
We suggest here that not only the isotropic energy but also the photon
index may depend on $\vartheta$. Another possibility is that
if short GRBs are the first 1~sec of the activity of the 
central engine, the spectrum in the early time might be
$\alpha_B\sim -0.58$ for both  the sub-jets in the core and the 
envelope. 
This is consistent with a high KS test probability for $E_p$  and 
$\alpha_B$ \citep{ggc03}. These possibilities may
have something to do with the origin of $\alpha_B\sim -1$ 
for the long GRBs.

\section{Discussions}
Let $\Delta\theta_{\rm sub}$, $\vartheta_c$ and $\bar{n}_s$ be the
typical opening half-angle of the sub-jet, the core size of the whole jet
and the mean multiplicity in the core. Then the total number of the
sub-jets ($N_\tot$) is estimated as
$N_\tot=\bar{n}_s(\vartheta_c/\Delta\theta_{\rm sub})^2\sim 10^3$,
so that the total energy of each sub-jet is $\sim 10^{48}$ erg. 
In our model, 
the event rate of long GRBs is in proportion to $\vartheta_c^2$. 
Let $M$ be the number of sub-jets in the 
envelope of the core with a  multiplicity $n_s=1$. 
Then the event rate of short GRBs is in proportion to
$M\Delta\theta_{\rm sub}^2$, so that $M\sim 10$ is enough to explain 
the event rate of short GRBs. 


Since the core may be regarded as a uniform jet,
our model for the XRFs is analogous to the off-axis uniform jet model
\citep{yin02,yin03b,yin04a}.
However the afterglow could have a different behavior between
the core-envelope sub-jet model and the uniform jet model.
In the uniform jet model, the afterglows of XRFs should resemble
the orphan afterglows that initially have a rising light curve
\citep{yin03a,g02}.
An orphan afterglow may be actually observed in XRF 030723 \citep{f04},
but the light curve may peak too early \citep{z03}.
The optical afterglow of XRF 020903 is not observed initially ($<0.9$ days)
but may not be consistent with the orphan afterglow \citep{so03}.
These problems could be overcome by introducing a Gaussian tail
with a high Lorentz factor around the uniform jet \citep{z03}
since the energy redistribution effects may bring the rising light curve
to earlier times \citep{z03,kg03}.
The afterglow of a short GRB is difficult to predict since it
could resemble both the orphan and normal afterglow depending on 
the sub-jet configuration within the envelope.

It is also found that our model can reproduce the bimodal
distribution of $T_{90}$ duration of GRBs observed by BATSE \citep{toma2004}.
In our model,
The duration of $n_s=1$ burst is determined by the angular
spreading time of one sub-jet emission, while
that of $n_s \geq 2$ burst is determined by the time interval between
the observed first pulse and the last one.
These two different time scales naturally lead a division of the burst
$T_{90}$ durations
into the short and long ones.
It has commonly been said that
the observed bimodal distribution of $T_{90}$ durations of BATSE bursts
shows the different origins of short and long GRBs.
However, the bimodal distribution is also available as a natural
consequence of
our unified model of short and long GRBs.
The clear prediction of our unified model is that
short GRBs should be associated with energetic SNe.
Indeed, one of short GRBs shows the possible association with a SN \citep{germany00}.
Even if the SNe are not identified with short GRBs due to
some observational reasons we predict that the spatial distribution of
short GRBs in host galaxies should be similar to that of the long GRBs.
Another prediction is that short GRBs should have the same total kinetic
energies as long GRBs, which might be confirmed by radio calorimetry.

Interestingly, our model has predicted short XRFs or short X-ray rich
GRBs \citep{yin04b}. 
They are observed when  isolated sub-jets are viewed slightly
off-axis.
The observed short XRF~040924 may be a kind of these bursts
\citep{fenimore04}.





\begin{thebibliography}{}




\bibitem[Fenimore et al.(2004)]{fenimore04}
Fenimore, E. et al. 2004, GCN Circ. 2735

\bibitem[Fynbo et al.(2004)]{f04}
Fynbo, J. P. U., Sollerman, J., Hjorth, J. et al.,
 2004, ApJ. 609, 962

\bibitem[Ghirlanda et al.(2004)]{ggc03}
Ghirlanda,~G., Ghisellini,~G., \& Celotti,~A.,
 2004, A\&A, 422, L55

\bibitem[Germany et al.(2000)]{germany00}
Germany, L. M., Reiss, D. J., Sadler, E. M. et al.,
2000, ApJ, 533, 320


\bibitem[Granot et al.(2002)]{g02}
Granot,~J., Panaitescu,~A., Kumar,~P., \& Woosley,~S.~E., 
2002, ApJ, 570, L61

\bibitem[Heise et al.(2001)]{He01a} 
Heise,~J., in 't Zand,~J., Kippen,~R.~M., \& Woods,~P.~M.,
2001, in Proc. Second Rome Workshop: Gamma-Ray Bursts in the Afterglow Era, 
ed. E.~Costa, F.~Frontera, \& J.~Hjorth (Berlin: Springer), 16 

\bibitem[Hjorth et al.(2003)]{hjorth03}
Hjorth, J.,  Sollerman, J.,  M\"{o}ller, P. et al.,
2003, Nature, 423, 847


\bibitem[Ioka \& Nakamura(2001)]{in01}
Ioka,~K., \& Nakamura,~T.,
2001, ApJ, 554, L163

\bibitem[Kippen et al.(2002)]{ki02} 
Kippen,~R.~M., Woods, P. M.,  Heise, J. et al. 
2002, AIP Conf. Proc. 662, 244 (astro-ph/0203114)

\bibitem[Kumar \& Granot(2003)]{kg03}
Kumar,~P., \& Granot,~J.,
2003, ApJ, 591, 1075

\bibitem[Nakamura(2000)]{nakamura2000}
Nakamura,~T.,
2000, ApJ, 534, L159

\bibitem[Sakamoto et al.(2004)]{s03}
Sakamoto,~T., Lamb, D. Q.,  Graziani, C. et al.,
 2004, ApJ, 602, 875

\bibitem[Soderberg et al.(2004)]{so03}
Soderberg,~A.~M., Kulkarni, S. R.,  Berger, E. et al.,
 2004, ApJ, 606, 994

\bibitem[Stanek et al.(2003)]{stanek2003}
Stanek, K.~Z.  Matheson, T., Garnavich, P. M. et al.,
2003, ApJ, 591, L17

\bibitem[Toma et al.(2004)]{toma2004}
Toma, K., Yamazaki, R., \& Nakamura, T.,
2004, astro-ph/0407012

\bibitem[Yamazaki et al.(2002)]{yin02}
Yamazaki,~R., Ioka,~K., \& Nakamura,~T.,
2002, ApJ, 571, L31

\bibitem[Yamazaki et al.(2003a)]{yin03a}
Yamazaki,~R., Ioka,~K., \& Nakamura,~T., 
2003a, ApJ, 591, 283

\bibitem[Yamazaki et al.(2003b)]{yin03b}
Yamazaki,~R., Ioka,~K., \& Nakamura,~T., 
2003b, ApJ, 593, 941

\bibitem[Yamazaki et al.(2003c)]{yyn03}
Yamazaki,~R., Yonetoku,~D., \& Nakamura,~T., 
2003c, ApJ, 594, L79

\bibitem[Yamazaki et al.(2004a)]{yin04a}
Yamazaki,~R., Ioka,~K., \& Nakamura,~T., 
2004a, ApJ, 606, L33

\bibitem[Yamazaki et al.(2004b)]{yin04b}
Yamazaki,~R., Ioka,~K., \& Nakamura,~T., 
2004b, ApJ, 607, L103

\bibitem[Zhang et al.(2004)]{z03}
Zhang,~B., Dai,~X., Lloyd-Ronning,~N.~M., \& 
M${\acute {\rm e}}$sz${\acute {\rm a}}$ros,~P.,
2004, ApJ, 601, L119




\end{thebibliography}
\end{document}